\documentclass[letterpaper, preprint, paper,11pt]{AAS}	

\usepackage[utf8]{inputenc}

\usepackage{graphicx}
\usepackage{amsmath}
\usepackage[version=4]{mhchem}
\usepackage{siunitx}
\usepackage{longtable,tabularx}
\usepackage{float}
\usepackage{placeins}
\usepackage{url}
\usepackage{hyperref}

\usepackage[utf8]{inputenc}
\usepackage{amssymb}
\usepackage{placeins}
\usepackage{bm}
\usepackage{siunitx}
\usepackage{graphicx}
\usepackage{subfigure}
\usepackage{placeins}
\usepackage{graphicx}
\usepackage{float}
\usepackage{footnpag}		
\usepackage{overcite}
\usepackage{bm}
\usepackage{amsmath}
\usepackage{subfigure}
\usepackage{caption}
\usepackage{algorithm}
\usepackage{algpseudocode}
\usepackage{footnpag}			      	
\usepackage{comment}

\PaperNumber{22-770} 

\title{Initial Orbit Determination for the CR3BP using Particle Swarm Optimization}

\author{David Zuehlke\thanks{Ph.D. Candidate, Space Technologies Lab, Embry-Riddle Aeronautical University, 1 Aerospace Blvd., Daytona Beach, FL, 32114.}, 
\  Taylor Yow\thanks{Master's Student, Space Technologies Lab, Embry-Riddle Aeronautical University, 1 Aerospace Blvd., Daytona Beach, FL, 32114.},
\ Daniel Posada\footnotemark[1], 
\ Joseph Nicolich \footnotemark[1]\thanks{Undergraduate Student, Space Technologies Lab, Embry-Riddle Aeronautical University, 1 Aerospace Blvd., Daytona Beach, FL, 32114.},
\ Christopher W. Hays \footnotemark[1],
\ Aryslan Malik\thanks{Visiting Professor, Space Technologies Lab, Aerospace Engineering Department, Embry-Riddle Aeronautical University, 1 Aerospace Blvd., Daytona Beach, FL, 32114.}
\ and Troy Henderson\thanks{Associate Professor, Space Technologies Lab, Aerospace Engineering Department, Embry-Riddle Aeronautical University, 1 Aerospace Blvd., Daytona Beach, FL, 32114.}
}

\begin{document}

\maketitle

\begin{abstract}
This work utilizes a particle swarm optimizer (PSO) for initial orbit determination for a chief and deputy scenario in the circular restricted three-body problem (CR3BP). The PSO is used to minimize the difference between actual and estimated observations and knowledge of the chief's position with known CR3BP dynamics to determine the deputy initial state. Convergence is achieved through limiting particle starting positions to feasible positions based on the known chief position, and sensor constraints. Parallel and GPU processing methods are used to improve computation time and provide an accurate initial state estimate for a variety of cislunar orbit geometries. 
\end{abstract}

\section{Introduction}
Increased interest in missions beyond the geosynchronous radius to cislunar (or XGEO) space necessitate viable methods of orbit estimation and space traffic management in this complex dynamical regime. Recent announcements by the Air Force Research Laboratory on the Cislunar Highway Patrol System (CHPS) solicit companies to present proposals for providing space traffic management in cislunar space.\cite{Erwin2022IndustryProposalsSoughtForCislunarHighwayPatrol-zo} Recent market analysis research claims that upwards of 250 lunar missions with a market value over \$100 Billion are expected by the year 2030.\cite{moon_market_analysis-wb} Given the interest by NASA and commercial entities greater understanding of satellite motion and tracking in the cislunar domain is needed. 

In the early 1960s, Szebehely compiled much of the available knowledge on the problem of restricted three-body orbits and published his findings as a reference for future development.\cite{Szebehely1967-mo} His work focused on the CR3BP, and provides a common framework for reference. Much work has been done studying the periodic orbits and studying their properties in the cislunar domain, including calculating initial conditions, transferring between orbits, and outlining the equations of motion to transition between ephemeris models.\cite{Gordon1991-qk,Wilson1993-hv,Gupta2020-sg,Grebow2006-ik,Pernicka1990-cb,Hiday1992-cj} Recent work by Greaves and Scheeres sought to lay out a framework using optical measurements alone for conducting cislunar space-situational awareness (SSA) under the assumption of CR3BP motion. They found that a single space based sensor placed in an orbit near the lunar L2 point could provide successful state estimation and maneuver detection for a variety of periodic orbit families such as near-rectilinear halo orbits (NRHO), and distant retrograde orbits (DRO). However, the optimal control based estimator required the inclusion of ``calculated'' angular rate measurements to stabilize the filtering estimates, and was limited to the CR3BP. \cite{Greaves_Scheeres_Daniel2021RelativeEstimationInTheCislunarRegime-jy,Greaves2021ObservationAndManeuverDetectionForCislunarVehicles-fb} Miller examined relative navigation for spacecraft in NRHOs, and used an Extended Kalman Filter (EKF) to estimate the relative states using a linearized model of the CR3BP and showed promising results.\cite{Miller2021_Relative_navigation_NRHO_EKF_NASA-wi} 

Further research in cislunar SSA includes work by Hall et. al. utilizing reachability set theory for detecting maneuvering objects in cislunar space.\cite{Hall2021_Reachability_based_approach_for_search_and_detection-li} The authors conducted extensive Monte Carlo trials for two distinct transfer orbits, one from L1 to L2, and an L2 to GEO maneuver. In both cases the maneuver was assumed to be bounded, and governed by CR3BP dynamics. LaFarge et. al. sought to leverage reinforcement learning for developing stationkeeping maneuver strategies and timings.\cite{LaFarge2021_Howell_Stationkeeping_in_multibody_orbits_leveragint_reinforcement_learning-tq} Once again though the analysis is limited to the CR3BP, a common theme among much of the current research on cislunar space.  
Khoury studied relative motion in the cislunar domain and outlined the relative and non-relative equations of motion for both the CR3BP and the ER3BP.\cite{Khoury2020Orbital_Rendezvous-lq} Further work by Greaves showed that optical observations were sufficient for simultaneous state estimation of both an observer and target spacecraft.\cite{Greaves_Scheeres_Daniel2021RelativeEstimationInTheCislunarRegime-jy} Fowler further studied the problem of cislunar Space Domain Awareness (SDA) and examined various observer placements ranging from earth-orbiting observatories to Lagrange point satellites and developed several metrics to aid in the creation of cislunar SDA constellation design.  

This work seeks to provide a novel method of cislunar initial orbit determination (IOD) using a numerical optimization approach. A deputy and chief satellite are simulated under CR3BP dynamics for a variety of observer and target orbit geometries. Then a particle swarm optimizer (PSO) is used to fit a set of observations (range, and angular, and angles only) to particle observations computed from propagating initial particle states forward to measurement times. Convergence of the PSO is assisted by including a constriction factor, initializing particles in a grid fashion, and limiting the scope of initial particle states. It is shown that the PSO converges to an accurate initial state estimate for the deputy satellite. Parallel processing and GPU processing methods are utilized to speed computation time. 
\section{Background}

\subsection{Three-Body Dynamics}
The three-body problem has been studied by mathematicians for more than 200 years, with significant developments coming in recent years with the increased interest in periodic orbits. The general three-body problem, though mathematically intriguing offers little in the way of practical applications. Simplifications such as the restricted three body problem allow for greater insight into the motion of a satellite in cislunar space without significant loss in accuracy.\cite{Szebehely1967-mo} The restricted three body problem's most general form is the Elliptical Restricted Three Body Problem (ER3BP) wherein the motion of the primary gravitational bodies is assumed to be elliptical. The ER3BP can be simplified further by assuming a circular orbit between the primary bodies, which describes the Circular Restricted Three Body Problem (CR3BP). Both the CR3BP and the ER3BP have been studied extensively for satellites orbiting in the earth-moon system.\cite{Szebehely1967-mo,Gordon1991-qk,Greaves_Scheeres_Daniel2021RelativeEstimationInTheCislunarRegime-jy,Franzini2020-vi,Galullo2022ClosedLoopGuidanceDuringCloseRangeRendezvousinaThreeBodyProblem-tf} In this study the dynamics model used will be the CR3BP, as much research has been done for investigating periodic orbits in the CR3BP.\cite{Howell1982-zl,LaFarge2021_Howell_Stationkeeping_in_multibody_orbits_leveragint_reinforcement_learning-tq} 

\begin{figure}[tbh!]
    \centerline{\includegraphics[scale=0.60]{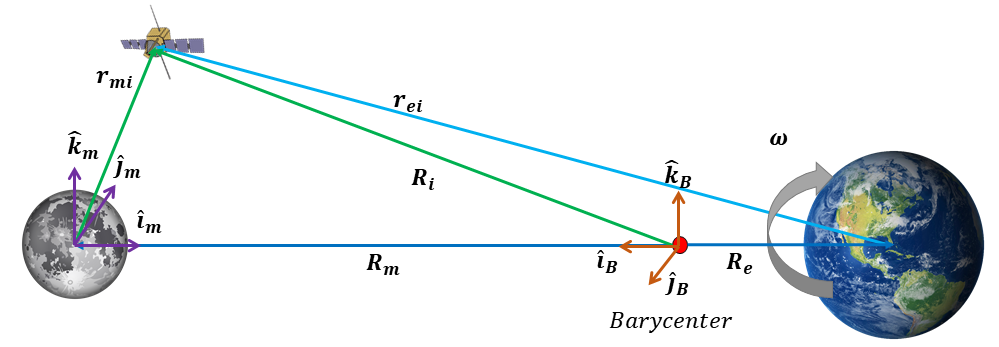}}
    \caption{Three-Body Coordinate Systems}
    \label{fig:3bp:coordinate:systems}
\end{figure}

The geometry of the CR3BP is shown in Fig. \ref{fig:3bp:coordinate:systems}. The coordinate frame chosen for this application is a rotating frame centered at the earth-moon barycenter, denoted the $B$ frame, and unit vectors given by $\hat{\mathbf{i}}_B$, $\hat{\mathbf{j}}_B$, and $\hat{\mathbf{k}}_B$. This is known as the barycentric synodic frame, since the rotation is aligned with the rotation of the moon about the earth. 

Note that in the CR3BP the earth-moon distance is constant and is equal to the moon's semimajor axis. The rotation rate is also constant and can be found from two-body relationships.\cite{vallado_mcclain_2013} The coordinate frame directions are defined with the $x-$axis pointing radially from the barycenter to the moon, the $z-$ axis in the earth-moon angular momentum direction, and the $y-$axis completes the right-handed set. The unit vectors are given by:
\begin{align}
    \hat{\mathbf{i}}_m & = \dfrac{\mathbf{r}_{em}}{r_{em}} \\
    \hat{\mathbf{j}}_m & = \hat{\mathbf{k}}_m \times \hat{\mathbf{i}}_m \\
    \hat{\mathbf{k}}_m & = \dfrac{\mathbf{r}_{em} \times \dot{\mathbf{r}}_{em}}{\left\| \mathbf{r}_{em} \times \dot{\mathbf{r}}_{em}\right\|}
\end{align}
Where $\mathbf{r}_{em}$ denotes the vector between the earth and the moon. The location of the barycenter can be found using the gravitational parameters of the earth $(\mu_e)$ and moon $(\mu_m)$, and the semi-major axis of the earth-moon system $(a)$. In the three-body problem, the combined, or non-dimensional gravitational parameter is of great importance and is given by: $\mu = \dfrac{\mu_m}{\mu_m + \mu_e}$. The combined gravitational power permits the scaling of the equations of motion into a non-dimensional form that is common in much of the literature on CR3BP orbits.\cite{Szebehely1967-mo}

The CR3BP equations of motion are shown in eqs. $($\ref{eqn:cr3bp:x}  $-$ \ref{eqn:cr3bp:z}$)$ and describe the motion of satellites in the barycentric-synodic $B-$ frame.\cite{Szebehely1967-mo,Franzini2020-vi} Figure \ref{fig:3bp:coordinate:systems} shows the geometric relationships of the barycentric synodic frame used to develop the CR3BP equations of motion. Note that the non-dimensional form of the equations are shown here in terms of the gravitational parameter $\mu$.

\begin{align}
    \ddot{x} &= 2 \dot{y} + x - \left(1 - \mu \right)\dfrac{x + \mu}{\left(\left(x + \mu\right)^2 + y^2 + z^2\right)^{3/2}} - \mu \dfrac{\left(x - \left(1 - \mu\right) \right)}{\left(\left(x - \left( 1 - \mu\right)\right)^2 + y^2 + z^2\right)^{3/2}} \label{eqn:cr3bp:x}\\
    \ddot{y} & =  -2 \dot{x} + y - \left(1 - \mu\right)\dfrac{y}{\left(\left(x + mu\right)^2 + y^2 + z^2\right)^{3/2}} -\mu \dfrac{y}{\left(\left(x - \left( 1 - \mu\right)\right)^2 + y^2 + z^2\right)^{3/2}} \label{eqn:cr3bp:y}\\
    \ddot{z} & =  -\left(1 - \mu\right)\dfrac{z}{\left(\left(x + \mu\right)^2 + y^2 + z^2\right)^{3/2}} - \mu \dfrac{z}{\left(\left(x - \left(1 - \mu\right)\right)^2 + y^2 + z^2 \right)^{3/2}}
    \label{eqn:cr3bp:z}
\end{align}

\subsection{Particle Swarm Optimization}
\begin{figure}[tbh!]
    \centerline{\includegraphics[scale=0.47]{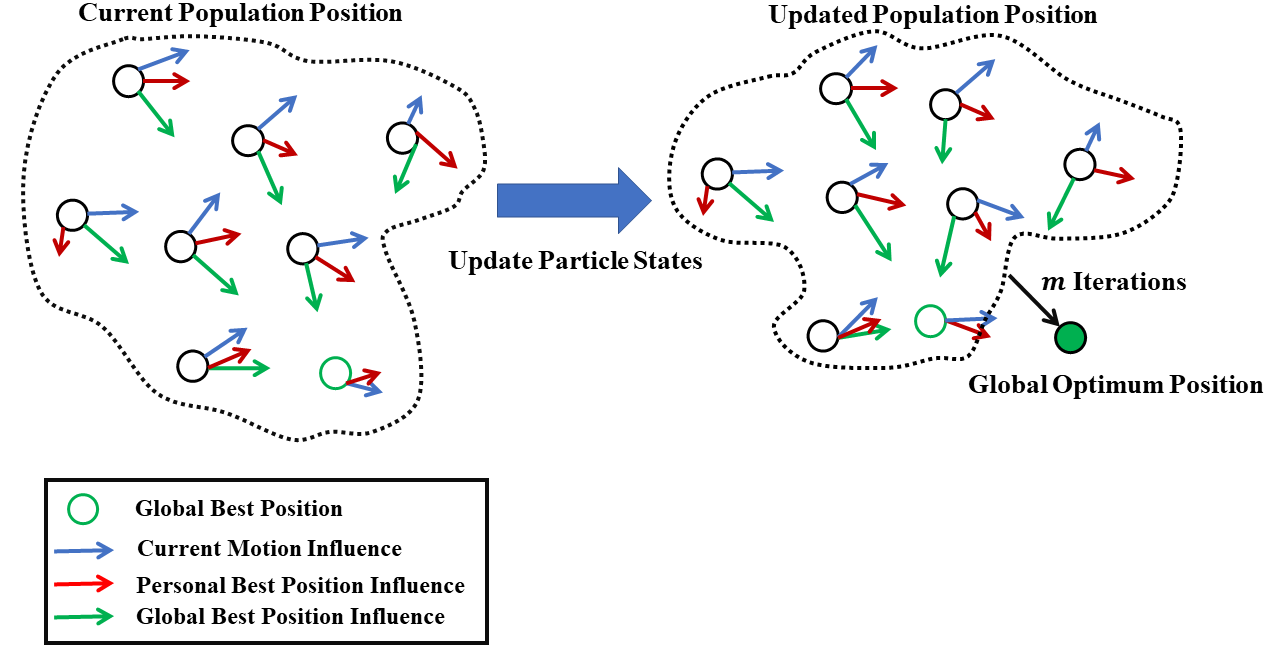}}
    \caption{PSO operation overview. Particle population moves under the influence of current best positions (blue arrows), current motion (red arrows), and current global best position (green arrows). After $m$ iterations the population converges to the global optimum solution.}
    \label{fig:pso:overview}
\end{figure}

One numerical, stochastic optimization method is the Particle Swarm Optimization (PSO) algorithm, which is inspired by the behavior of birds and takes advantage of information sharing among the swarm, called ``collective intelligence” \cite{Kennedy,Poli,Clerc}. The PSO algorithm is metaheuristic in nature, straightforward to configure, and provides the capacity to efficiently converge on an optimal solution. The PSO is a population-based algorithm where each particle element in the population has a $N$-dimensional position representing potential state value solutions. Each particle also has an associated cost value, and a velocity which determines the position update. In this way, the particles ``swarm" to the state that produces the global minimum cost value. The PSO benefits from the social interaction of a large number of particles. Thus, it is a global optimization technique that can provide solutions within a broad search space, but not to a high degree of precision without a significant amount of processing. The basic idea of how a PSO functions is shown in Fig. \ref{fig:pso:overview}. The population moves under the influence of each particle's ``best'' positions, the populations ``best'' position, and particle's current motion. These factors combined with weighting factors determine particle updates to the $N$ dimensional state. 
The particle motion weighting is influenced by cognitive and social hyper-parameters (denoted $\eta_1$ and $\eta_2$ respectively) which determine the influence of a particle's individual position compared to the swarm's current global best position. The PSO implementation is further enhanced by the addition of a constriction factor $\kappa$, which minimizes the number of iterations necessary to attain the desired accuracy, hence benefiting the method by reducing computing time \cite{malik2021constriction}. The constriction factor is a function of social and cognitive hyper-parameters that maintains an optimal balance between exploration and exploitation.
\begin{equation}
    \kappa=\frac{2}{\lvert 2-\phi-\sqrt{\phi^2-4\phi} \rvert},\;\phi=\eta_1+\eta_2>4
    \label{eqn:constriction:factor}
\end{equation}

It was also demonstrated that the initialization of the particles in the solution space has a significant impact on the speed of the convergence \cite{malik2021constriction,malik2021gridinitial}. A uniform ``grid-like'' initialization can be combined with constraints on the position and velocity of the particle in order to further shorten the computation time which will be explored in this work. 
\section{Methodology} 
\subsection{Deputy Chief Scenario for Cislunar IOD}
\begin{figure}[tbh!]
    \centerline{\includegraphics[scale=0.47]{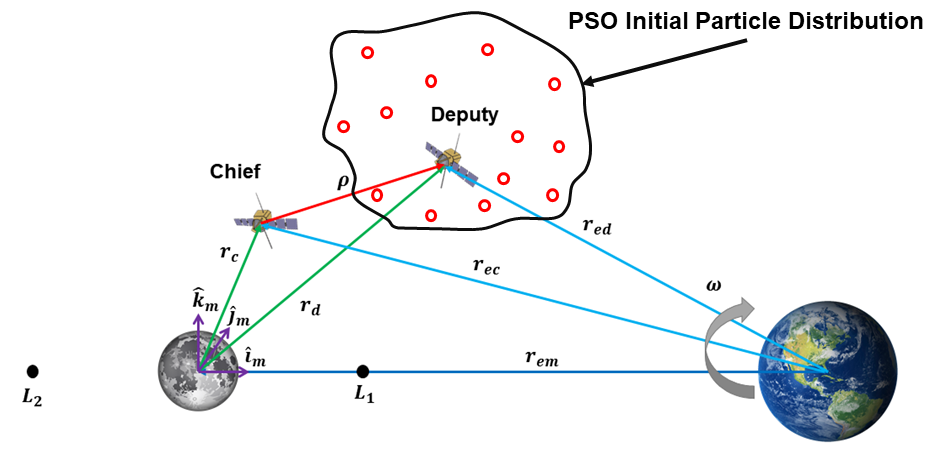}}
    \caption{Deputy chief initial orbit determination scenario. PSO initial states are seeded around a probable deputy position (shown as red circles).}
    \label{fig:pso:deputy:chief}
\end{figure}

The scenario of interest is a deputy chief scenario in cislunar space. The chief's states are assumed to be known and available for the orbit determination process. The geometry of the scenario is shown in Fig. \ref{fig:pso:deputy:chief}. The chief is assumed to be able to capture range, and angular measurements. 
In order for the PSO to work, an initial population of possible deputy states are required, and are seeded around the probable guess, shown by the red circles in Fig. \ref{fig:pso:deputy:chief}. Trajectories for the deputy and chief satellites are generated by propagation via the CR3BP non-dimensional equations of motion. Periodic orbits are of the most interest and initial conditions are defined from NASA's JPL Horizons tool which has a database of periodic orbits in the CR3BP.\cite{JPL_horizons-yx}

\subsection{Particle Swarm Optimizer}
Algorithm 1 outlines the Particle Swarm Optimization method used. Where the global minimum $\hat{\mathbf{g}}$ is defined by  the particle state with the minimum cost. Particle states $\mathbf{x}_i$ define each initial guess for the deputy satellites state (both position and velocity). Particle velocities $\mathbf{v}_i$ determine the movement within the six dimensional search space for an initial state. Hyper-parameters that can be tuned are the particle inertia $(\omega)$ which typically lies in the range $0.4 < \omega < 1.4$ and controls the speed of particle velocity updates. The exploration factor, or self-confidence $\eta_1$ that defines particle confidence in it's own solution. The swarm-confidence or exploitation factor $\eta_2$, that determines particle confidence in the current global best solution. 

\begin{algorithm}[htb!] 
\caption{Particle Swarm Optimization Algorithm Pseudo-code}\label{alg:cap}
\begin{algorithmic}[1]
\label{alg:particle:swarm}
\State Initialize particle states: $\mathbf{x}_i$ and $\mathbf{v}_i$ for $i = 1, . . ., m$
\State $\hat{\mathbf{x}}_i \gets \mathbf{x}_i$ and $\hat{\mathbf{g}} = \textit{min}$  $J(\mathbf{x}_i)$ for $i = 1,...,m$
\For{$n = 1$ to max iterations $N$ \do} 
\For{$i = 1$ to number particles $m$ \do}
\State $J_i \gets J(\mathbf{x}_i)$ Find current cost of particle: 
\State Check if personal best cost:
\If{$J_i < J_{best_i}$} 
    \State $J_{best_i} \gets J_i$ 
    \State$\mathbf{x}_{best_i} \gets \mathbf{x}_i$ Update personal best state
    \EndIf
\State Randomly generate $r_1$, $r_2 \in U\left[0, 1\right]$ for particle velocity update:
\State Update particle velocity: $\mathbf{v}_i \gets 
\omega \mathbf{v}_i + \eta_1 r_1 \left(\mathbf{x}_{best_i} - \mathbf{x}_i\right) + \eta_2 r_2 \left(\hat{\mathbf{g}} - \mathbf{x}_i \right)$
\State Update particle position: $\mathbf{x_i} \gets \mathbf{x}_i + \mathbf{v}_i$
\EndFor
\State Check if new global best:
\If{$min(J_i\left(\mathbf{x}_i\right)) < J\left(\hat{\mathbf{g}}_i\right) $}
\State $\hat{\mathbf{g}}_i \gets \mathbf{x}_i$
\EndIf
\State Check if tolerance met for convergence
\If{$g_{error} <$ $\tau$ (Algorithm 2)}
    \State break
\EndIf
\EndFor
\end{algorithmic}
\end{algorithm}
 The heart of the PSO algorithm for computing an initial state for the deputy spacecraft comes from the cost function $J(\mathbf{x}_i)$. The cost function minimizes the difference between actual and computed measurements for each of the particles. Measurements are taken to be range, and azimuth and elevation angular measurements, denoted $\rho, \alpha,$ and  $\beta$ respectively. Particle states consist of a six-dimensional state vector containing a possible deputy initial position and velocity $\mathbf{x}_i =\begin{bmatrix} \mathbf{X}_0^T & \mathbf{V}_0^T\end{bmatrix}$. Thus each particle position is given by:  $\mathbf{x}_i \in \mathbb{R}^6$. Each particle state is then propagated forward in time using the CR3BP equations of motion (eqs. \ref{eqn:cr3bp:x} - \ref{eqn:cr3bp:z}) to all measurement times $t_k$. Next the relative position vector from the known chief position to the propagated particle deputy position is calculated as well as the line of sight (LOS) vector.
 \begin{equation}
     \boldsymbol{\rho}(t_k) =  \mathbf{r}_{d_i}(t_k) - \mathbf{r}_c(t_k) 
     \label{eqn:rel:pos}
 \end{equation}
 \begin{equation}
     \mathbf{L}_{t_k} = \dfrac{\boldsymbol{\rho}(t_k)}{\|\boldsymbol{\rho}(t_k)\|}
 \end{equation}
 where $\mathbf{r}_{d_i}(t_k)$ denotes the position vector of the $i$th particle at time $t_k$. Measurements are then computed for each timestep as range and azimuth, elevation angles. The range is simply the norm of the relative position vector $\rho(t_k) = \|\boldsymbol{\rho}(t_k)\|$. And the angular measurements are calculated from the LOS vector components as: 
 \begin{align}
     \alpha(t_k) &= \textit{atan2}\left(\dfrac{\mathbf{L}_{t_k}(2)}{\mathbf{L}_{t_k}(1)}\right) \\
     \beta(t_k) & = \arcsin{\left(\mathbf{L}_{t_k}(3)\right)}
 \end{align}
 where the parenthetic argument $(1,2,3)$ denotes the $x,y,z$ component of the line of sight vector $\mathbf{L}_{t_k}$ respectively. With all the elements defined, the measurement function is given by equation \ref{eqn:meas:function}.
 \begin{equation}
     \mathbf{y}(t_k) = \begin{bmatrix} \rho(t_k) \\
     \alpha(t_k) \\
     \beta(t_k)
     \end{bmatrix}
     \label{eqn:meas:function}
 \end{equation}
 For each time step the measurement residuals $\mathbf{b}_{t_k}$ are computed as the difference from the true measurement $\Tilde{y}(t_k)$ and the particle predicted measurement $y(t_k)$ as:
 \begin{equation}
     \mathbf{b}(t_k) = \Tilde{y}(t_k) - y(t_k)
 \end{equation}
The measurement residuals for each time step are then squared and weighted by the expected sensor noise level and formed into a row vector. Where measurement weights are defined as the inverse square of the sensor expected noise value $w_j = \dfrac{1}{\sigma_j^2} $ where $j = 1,2,3$ denotes range, azimuth, and elevation noise levels. Thus a row vector of residuals is formed as shown in eq. \ref{eqn:b:total}
\begin{align}
    \mathbf{b}_{total} = & \left[\mathbf{b}(t_1)^T W \mathbf{b}(t_1) \ \ \ \mathbf{b}(t_2)^T W \mathbf{b}(t_2) \ \ . . . \ \  \mathbf{b}(t_k)^T W \mathbf{b}(t_k) \right] \label{eqn:b:total}\\ 
    W = & \begin{bmatrix} w_1 & 0 & 0 \\
    0 & w_2 & 0 \\
    0 & 0 & w_3
    \end{bmatrix}
\end{align}

Once all measurement residuals are formed into a row vector, the final scalar cost from the current particle is calculated as the the square of the row vector as:
\begin{equation}
    J = \mathbf{b}_{total}^T \mathbf{b}_{total}
    \label{eqn:total:cost}
\end{equation}
Once the cost is calculated for a given particle, the current cost is compared to the particles personal best (lowest) cost. If the current cost is lower than the personal best, then the particle's personal best state is updated and is used for the update in the particle states as outlined in algorithm \ref{alg:cap}.
\subsection{Ending Conditions}
The implementation of a desired error tolerance into the PSO offers a reduction in run time. The optimization process can be terminated at a point where allowing the optimizer to continue propagating would significantly prolong run time while providing very little improvement in the global best state error. By tracking global cost as it trends downward, the error between current and former iterations' global cost can be utilized to impose a tolerance. Due to the random search nature of PSO, though, as particles converge on a global best, one particle may remain "best" for several iterations until succeeded by another particle which has found a lower cost. This results in a downward staircase-like trend which causes the global cost function to occasionally produce a constant value for more than one iteration at a time. Hence, it is sometimes impractical to compare the error between a current global cost and its former value alone, as the staircase profile would almost always cause the run to terminate prematurely. However, because global cost never increases, a span of global costs can be analyzed to determine the error between them. Experimentation on the necessary length of this span eventually arrived at using a vector of three global costs, as the global cost sometimes remained constant for two iterations but rarely remained constant for greater than three iterations. This value resulted in global costs below $10^{-2}$ while still terminating the run prior to reaching the maximum specified quantity of iterations, thus improving run time. For the sake of this paper, however, the provided samples were allowed to run for the entire duration of the specified iteration count. Future work can further investigate optimizing the number of iterations required to consistently achieve results within a certain precision and number of iterations by better refining the tolerance condition algorithm. 

\begin{algorithm} 
\caption{Tolerance Algorithm Pseudo-code}\label{alg:cap}
\begin{algorithmic}[1]
\label{alg:tolerance}
\State Initialize global best matrix 

\For{$i = 1$ to max iterations $N$ \do} 
\State Append current global best to global best matrix
\If{$i > 10$ minimum runs for algorithm \do}
\State Determine error between current global cost and global cost 10 iterations prior:
\State {$K = \left | (cost_i-cost_{i-10})/cost_i \right |$}
\EndIf
\If{$K<\tau$ is met \do}
\State{Run terminates}
\Else{ $\tau$ not met\do} 
\State Run continues
\EndIf
\EndFor
\end{algorithmic}
\end{algorithm}

\subsection{Local Minimization}
The PSO is a metahueristic global optimizer and is very good at discovering minimum in a global sense. For mutli-minimum problems however, particles can get ``stuck'' in a local minimum. A common method to increase accuracy is to use a local minimizer after a global optimization has been run to get the states as close as possible to the desired states. 
The method chosen for local minimization is non-linear least squares using the MATLAB implementation of the Levenberg-Marquardt (LM) method.\cite{Kenneth_Levenberg1944-qu,Marquardt1963-jj,More1978-ji} The non-linear least-squares problem is setup to minimize the difference between true and predicted measurements from the estimated state of the deputy satellite. The LM method seeks to minimize a function of the form: 
\begin{equation}
    min_x f(\mathbf{x}) = \|\mathbf{F}(\mathbf{x})\|_2^2 = \sum_i^N F_i^2(\mathbf{x})
    \label{eqn:non:linear:lsq}
\end{equation}
Where the vector $\mathbf{F}$ is given as the residuals for a set of measurements, which are subsequently squared in order to minimize the error between the truth measurements and the estimated measurements.  The objective function is a slightly modified version of the cost function employed for the PSO, where the residuals are formed as as a row vector in the following form.
\begin{equation}
    \mathbf{F} = \begin{bmatrix} J_1 & J_2 & ... & J_k
    \end{bmatrix}
    \label{eqn:nonlinlsq:cost}
\end{equation}
The elements of $\mathbf{F}$  are given by a slightly modified version of the cost function employed for the PSO. Where instead of summing all measurement residuals and then computing the weighted square, each measurement residual is computed and squared with the measurement weights. Each individual cost element then takes the form:
\begin{equation}
    J_k = \mathbf{b}^T(t_k) W \mathbf{b}(t_k)
\end{equation}
Where $\mathbf{b}(t_k) = \Tilde{y}(t_k) - y(t_k)$ is the measurement residual at time $t_k$. The output of the non-linear least squares minimization is an initial state for the deputy that has now moved closer to the truth than the result of the PSO. Simulation results showed that using the non-linear optimization after running the PSO was shown to reduce the final cost significantly and the error in the IOD estimate. 

\subsection{Analysis of Local Minimization} 
Further analysis into non-linear least squares aimed to identify possible trends which could allow local minimization to occur earlier on in the cost reduction process. If the PSO could reach a point where, regardless of further iterations, the non-linear least squares would continue finding the same minimum, then theoretically a high-precision local error could be found with further reduced run time. To visualize non-linear least square's performance through all of PSO's iterations, the local minimization algorithm was executed at each corresponding state error, then plotted alongside the original PSO error reduction for that case sample. It was found that while non-linear least squares continued oscillating as the PSO error decreased, it consistently offered a substantial reduction in error for a moderate number of iterations, until PSO propagated so many times that the local minimizer offered no remarkable benefit. 

\begin{figure}[tbh!]
    \centerline{\includegraphics[scale=0.47]{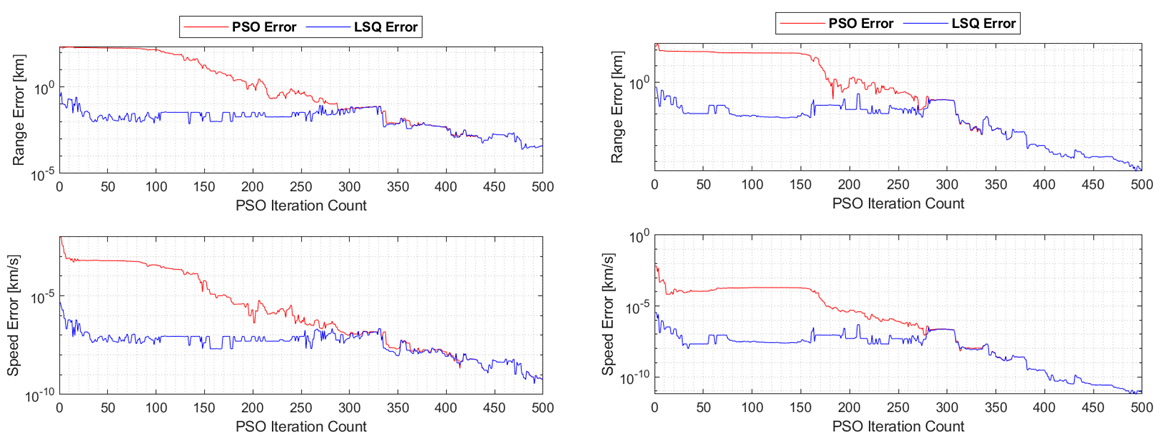}}
    \caption{Visualization of LSQ error reduction per iteration given corresponding PSO state error,
    300 particles (left) and 500 particles (right).}
    \label{fig:lsq:nonlin_100_particles}
\end{figure}

\subsection{PSO Optimization}
One of the main challenges to perform Particle Swarm Optimization on the CR3BP is how to accelerate the computation time and the use of more particles to improve the error and convergence. The PSO algorithm was coded using MATLAB's parallel computing toolbox to improve run-time and code efficiency. Currently multiple calculations such as population spawn and steps per particle are processed in CPU cores.
GPUs have quickly emerged as inexpensive parallel processors due to their high computation power. There are different CUDA libraries optimized to work with the hardware than can efficiently reduce the computation time. Future work will include the portability of the code to open source using python, vectorization to minimize loops, and implementation of CUDA core acceleration using gpuArrays with libraries such as pyCuda. \cite{vanneschi2011comparative,lalwani2019survey}.

\section{Results}
Results are shown for running the IOD method for the CR3BP on two scenarios for deputy and chief satellites operating on L2 periodic orbits of both HALO and axial kinds. The initial particle positions and velocities for the deputy were bounded by a Gaussian sphere of 250 km and 0.1 km/s respectively. Initial particles were randomly seeded around the true deputy position using MATLAB's builtin $\textit{rand}$ function and scaled appropriately by the limits in position and velocity. 

The results of both scenarios are summarized in Tab. \ref{tab:summary:results}. Both scenarios had the deputy in an L2 southern HALO orbit with a period of 7.15 days. The range error for scenario 1, was sub-kilometer for the PSO results. Note that the non-linear least squares (NLSQ) local optimizer was able to achieve near meter level accuracy for range and sub $m/s$ speed accuracy. 
\begin{table}[h!]
\caption{\label{tab:summary:results} Summary of PSO and nonlinear least squares (NLSQ) results. In both cases NLSQ produced a lower error after PSO terminated.}
\vspace{-0.4cm}
\begin{center}
\begin{tabular}{|l|ll|ll|}
\hline
\textbf{}                         & \multicolumn{2}{l|}{\textbf{Scenario   1}}        & \multicolumn{2}{l|}{\textbf{Scenario   2}}        \\ \hline
\textbf{}                         & \multicolumn{1}{l|}{\textbf{PSO}} & \textbf{NLSQ} & \multicolumn{1}{l|}{\textbf{PSO}} & \textbf{NLSQ} \\ \hline
\textbf{Range Error   {[}km{]}}   & \multicolumn{1}{l|}{0.3276}       & 0.0304        & \multicolumn{1}{l|}{6.072}  & 0.0177   \\ \hline
\textbf{Speed Error   {[}km/s{]}} & \multicolumn{1}{l|}{2.84E-06}     & 3.40E-08      & \multicolumn{1}{l|}{3.27E-05}     & 5.77E-08      \\ \hline
\textbf{Minimum Cost}             & \multicolumn{1}{l|}{0.0493}       & 8.56E-11      & \multicolumn{1}{l|}{1.667}        & 7.34E-11      \\ \hline
\end{tabular}
\end{center}
\end{table}


\subsection{Scenario 1}
Scenario 1 is a cislunar SDA scenario where the deputy is in a HALO orbit with a period of 7.15 days, and the chief is placed in a HALO orbit with a period of 13.8-days. The initial conditions used for the scenario are shown in Table \ref{tab:scenario1}. The scenario was propagated for 7 days, and a total of 35 measurements were taken, equally spaced through the total simulation time-span. 

\begin{figure}[tbh!]
    \centerline{\includegraphics[scale=0.49]{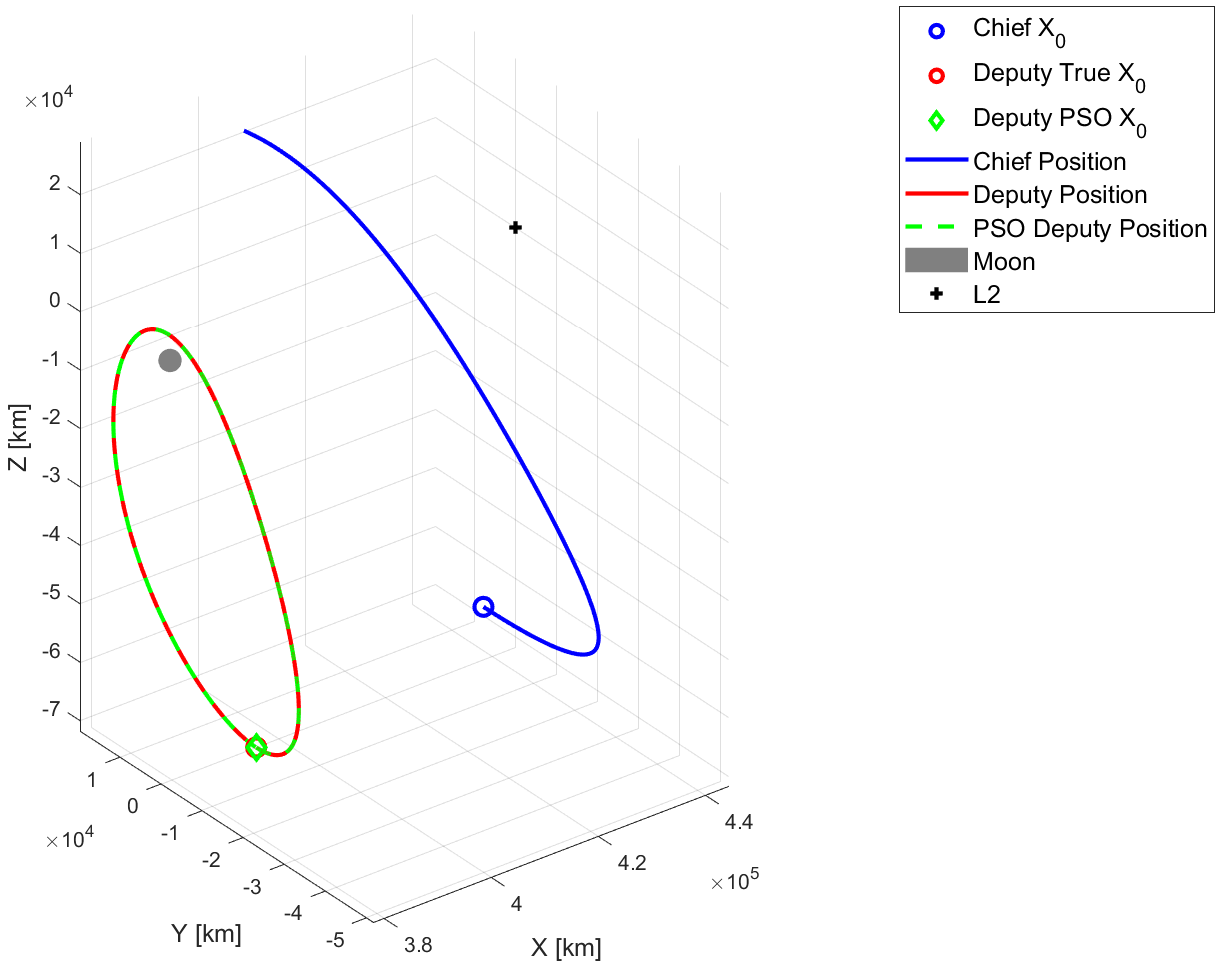}}
    \caption{Scenario 1: Chief and deputy HALO orbits depicted over 7-day period with PSO deputy orbit overlay.}
    \label{fig:scenario1:PSO:optimizer:results}
\end{figure}

\begin{table}[h!]
\caption{\label{tab:scenario1} Scenario 1 initial conditions. The scenario was propagated for 7 days and utilized 35 measurements.}
\vspace{-0.5cm}
\begin{center}
\begin{tabular}{|l|l|l|}
\hline
                                     & \textbf{Deputy}       & \textbf{Chief}      \\ \hline
\textbf{$x_0$ (LU)}                     & 1.140135389           & 1.029726968         \\ \hline
\textbf{$y_0$ (LU)}                     & 0                     & 0                   \\ \hline
\textbf{$z_0$ (LU)}                     & -1.63176653574390E-01 & -1.869397163946E-01 \\ \hline
\textbf{$v_{x_0}$   (LU/TU)}               & 6.13321115086310E-15  & -5.585615805585E-14 \\ \hline
\textbf{$v_{y_0}$   (LU/TU)}               & -0.223383154          & -0.119441863        \\ \hline
\textbf{$v_{z_0}$   (LU/TU)}               & 1.78644826151404E-15  & -9.803996218373E-13 \\ \hline
\textbf{Jacobi   constant (LU2/TU2)} & 3.06                  & 3.04                \\ \hline
\textbf{Period   (days)}             & 13.8                  & 7.15                \\ \hline
\end{tabular}
\end{center}
\end{table}
Figure \ref{fig:scenario1:PSO:optimizer:results} shows the scenario propagated for 7-days. The chief orbit is shown, and the deputy orbit and the orbit computed by the PSO is also plotted. Note the close agreement and the position overlapping of the PSO and truth comparison positions.

\begin{figure}[tbh!]
    \centerline{\includegraphics[scale=0.49]{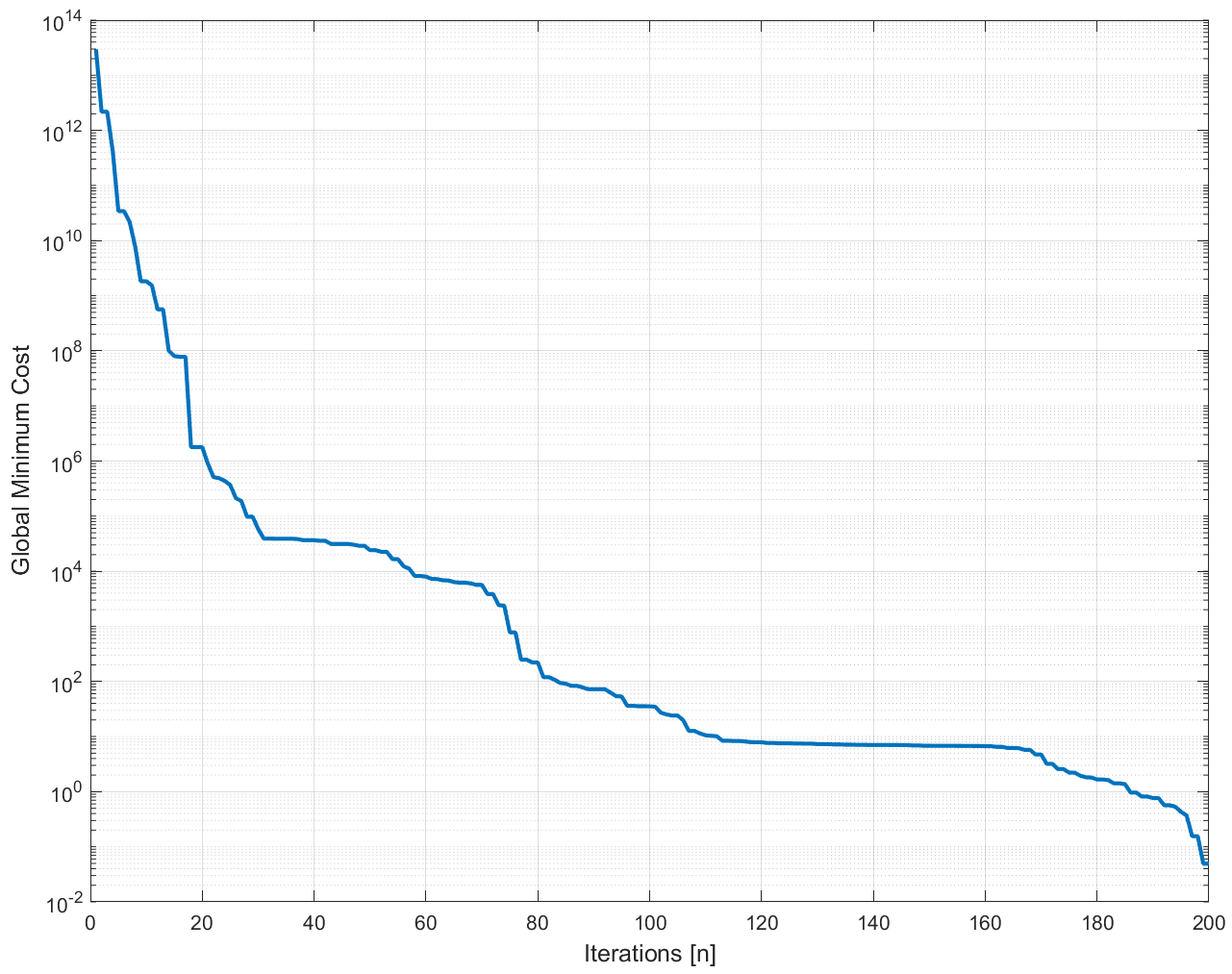}}
    \caption{Scenario 1: Global cost per iteration, decreasing until iteration counter is reached or tolerance is met.}
    \label{fig:scenario1:minimum:cost}
\end{figure}

Figure \ref{fig:scenario1:minimum:cost} shows the cost value for all iterations that the PSO ran through. Note the distinctive extended staircase pattern as the optimizer seeks the global minimum, and that the maximum number of iterations was hit, indicating that the tolerance was not met. 

\begin{figure}[tbh!]
    \centerline{\includegraphics[scale=0.49]{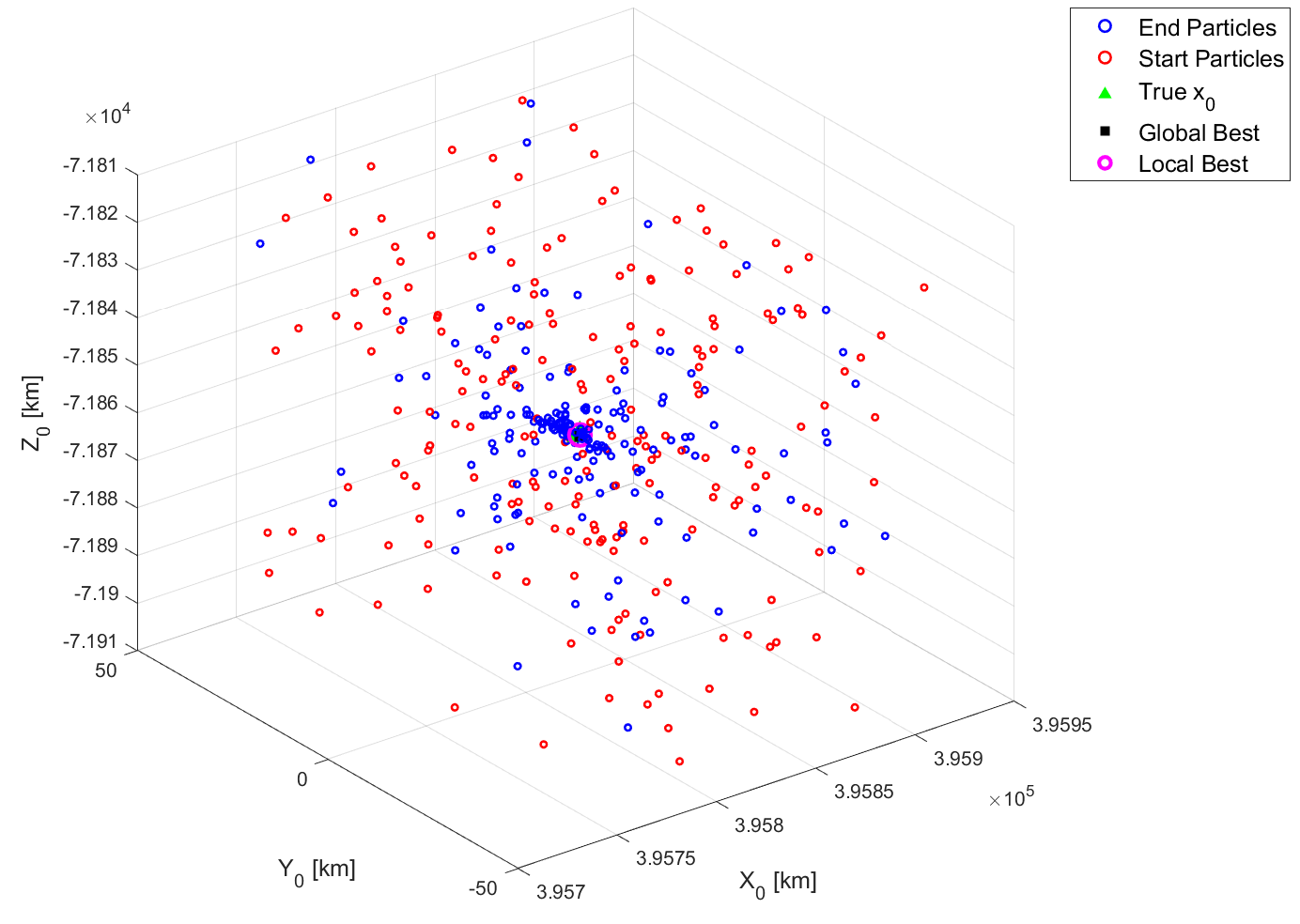}}
    \caption{Scenario 1: Initial (red) and final state (blue) particle positions in three axes, swarming to global best.}
    \label{fig:scenario1:particle:positions}
\end{figure}

\begin{figure}[tbh!]
    \centerline{\includegraphics[scale=0.49]{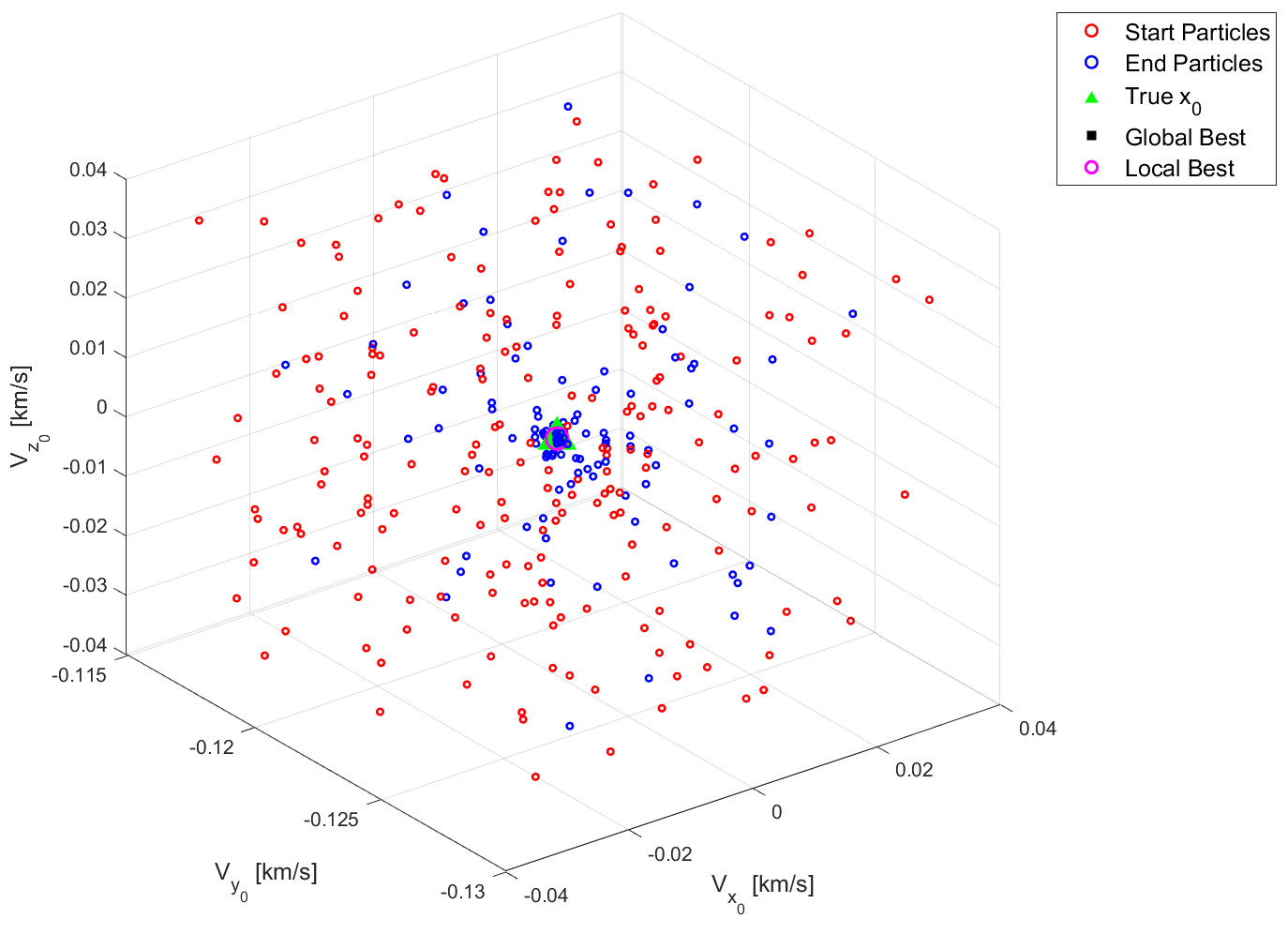}}
    \caption{Scenario 1: Initial (red) and final state (blue) of particle velocity in three axes.}
    \label{fig:scenario1:particle:velocities}
\end{figure}

Figures \ref{fig:scenario1:particle:positions} and \ref{fig:scenario1:particle:velocities} show the initial distribution of particles in position and velocity. Note that a majority of particles converge around the true initial position and velocity states. And that the local minimizer lies extremely close to the true state. 

\begin{figure}[tbh!]
    \centerline{\includegraphics[scale=0.49]{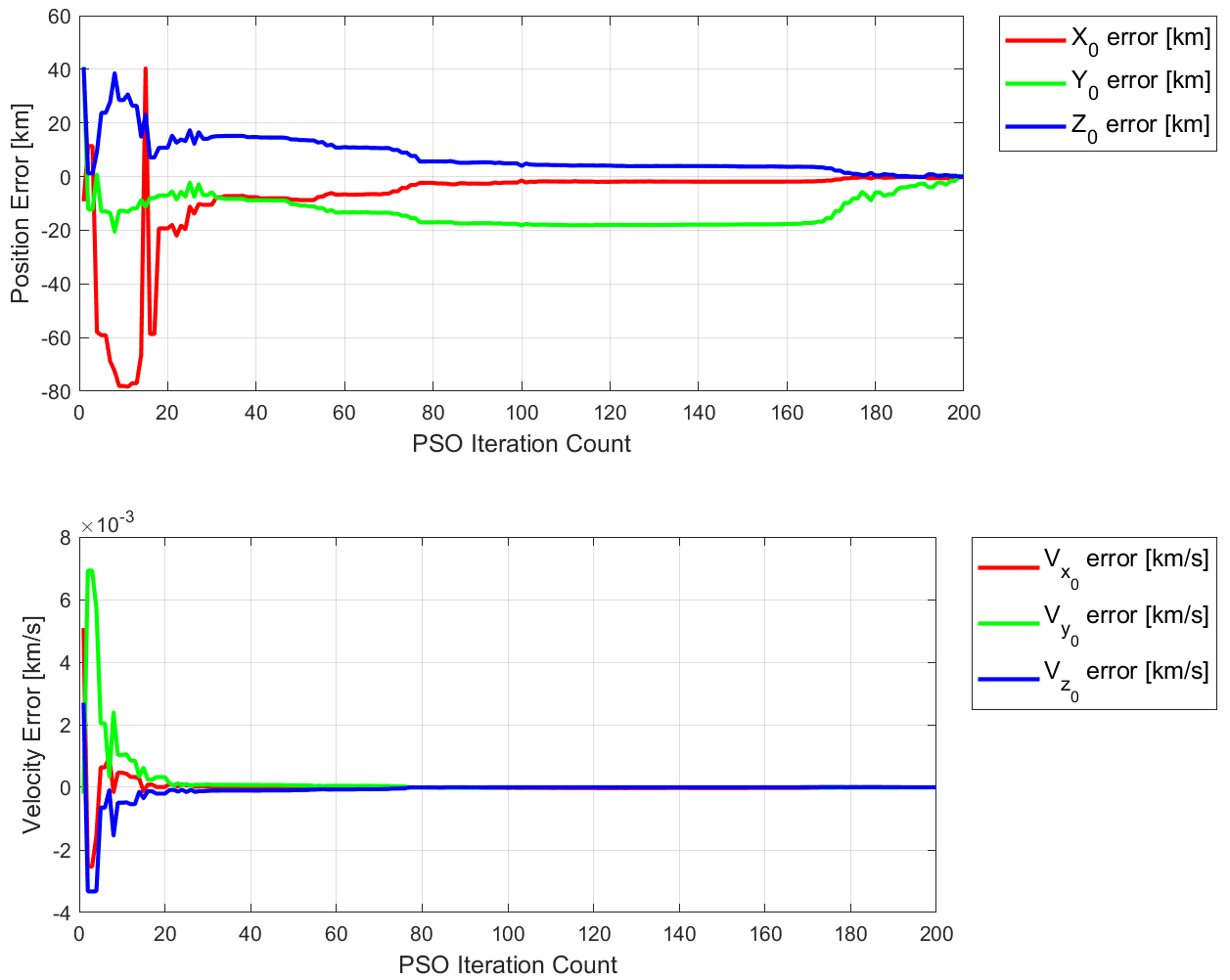}}
    \caption{Scenario 1: Position error in each axis per iteration (top) and velocity error in each axis per iteration (bottom).}
    \label{fig:scenario1:position:velocity:error}
\end{figure}

\begin{figure}[tbh!]
    \centerline{\includegraphics[scale=0.49]{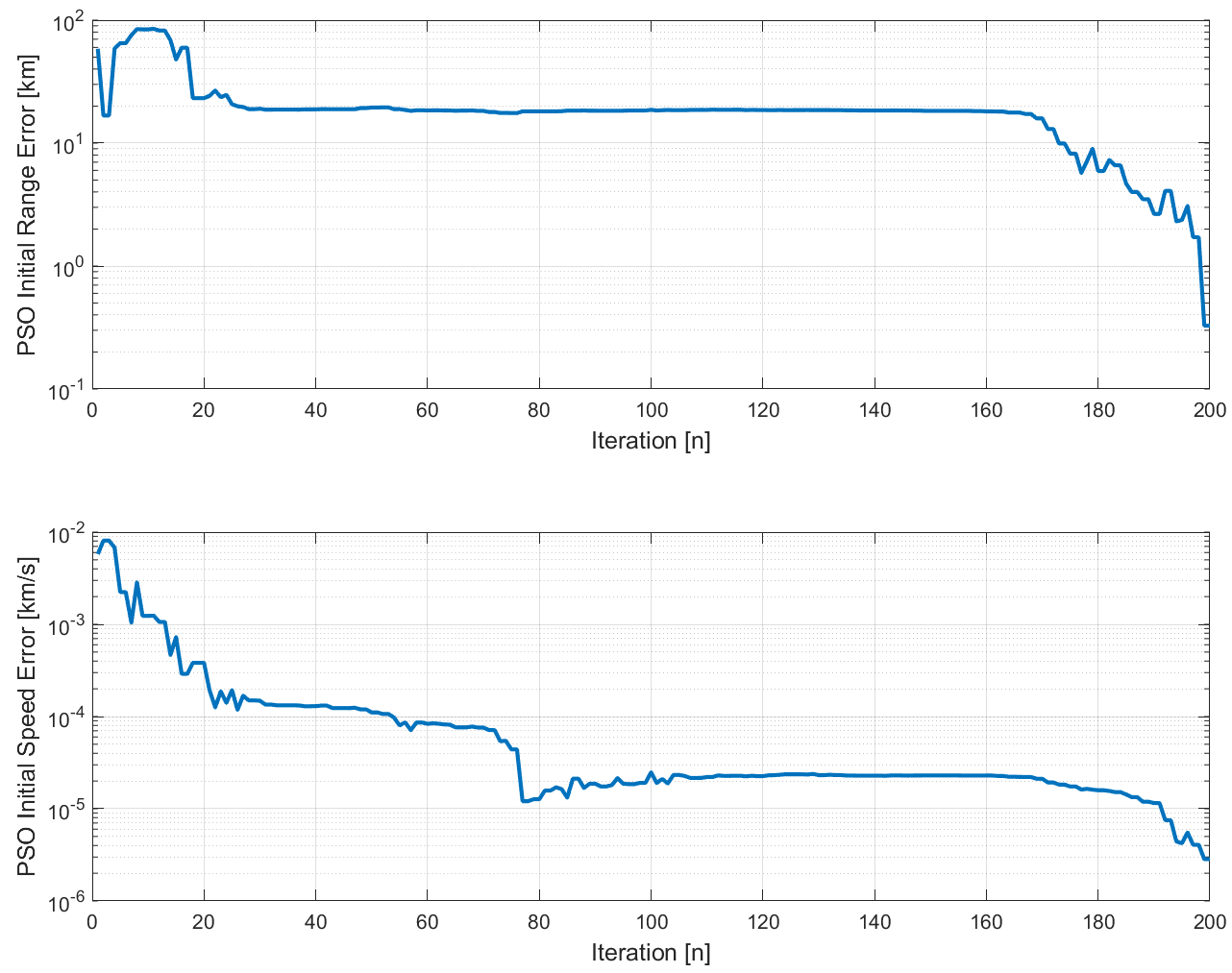}}
    \caption{Scenario 1: Magnitude of position error per iteration (top) and magnitude of velocity error per iteration (bottom).}
    \label{fig:scenario1:range:speed:error}
\end{figure}

Figure \ref{fig:scenario1:position:velocity:error} shows the position and velocity error of deputy initial conditions as the PSO converges towards the truth. Note that for the first few iterations there are large fluctuations in the error before a general decreasing trend is observed. The errors in position are summarily larger than the errors in velocity. Similarly, Fig. \ref{fig:scenario1:range:speed:error} shows the range and speed error of the initial state estimate as the PSO converges.  

Overall it can be seen that the PSO was able to determine a relatively close state estimate, and the local non-linear least squares optimizer was able to further refine that state error to within a few meters in position and sub-meter per second in velocity. 

\subsection{Scenario 2}
The second scenario is another cislunar SDA type application. The deputy is in the same orbit as scenario 1, but this time the chief is in an L2 Axial orbit with a period of 19.1 days. Much sparser measurements were utilized, with a total of 10 measurements taken over the 7 day propagation period. The PSO and NLSQ errors were both very low, with the PSO achieving a rang error $< 7 km$ and a speed error $< 0.05 m/s$. The initial particles for PSO were seeded with error bounds of $250km$ in position and $0.1 km/s$ in velocity from the true deputy position. Overall results are very similar to scenario 1, and follow the same trends. 

\begin{table}[h!]
\caption{\label{tab:scenario2} Scenario 2 initial conditions. The scenario was propagated for 7 days and utilized 10 measurements.}
\begin{tabular}{|l|l|l|}
\hline
\textbf{}                            & \textbf{Deputy}       & \textbf{Chief}        \\ \hline
\textbf{$x_0$ (LU)}                     & 1.140135389           & 1.21996614837886E+00  \\ \hline
\textbf{$y_0$ (LU)}                     & 0                     & 0                     \\ \hline
\textbf{$z_0$ (LU)}                     & -1.63176653574390E-01 & -2.49454877925228E-16 \\ \hline
\textbf{$v_{x_0}$   (LU/TU)}               & 6.13321115086310E-15  & 1.10859609358602E-15  \\ \hline
\textbf{$v_{y_0}$   (LU/TU)}               & -0.223383154          & -4.27475884185211E-01 \\ \hline
\textbf{$v_{z_0}$   (LU/TU)}               & 1.78644826151404E-15  & 4.09809301627323E-03  \\ \hline
\textbf{Jacobi   constant (LU2/TU2)} & 3.06                  & 3.01E+00              \\ \hline
\textbf{Period   (days)}             & 13.8                  & 19.1                  \\ \hline
\end{tabular}
\end{table}

\begin{figure}[tbh!]
    \centerline{\includegraphics[scale=0.49]{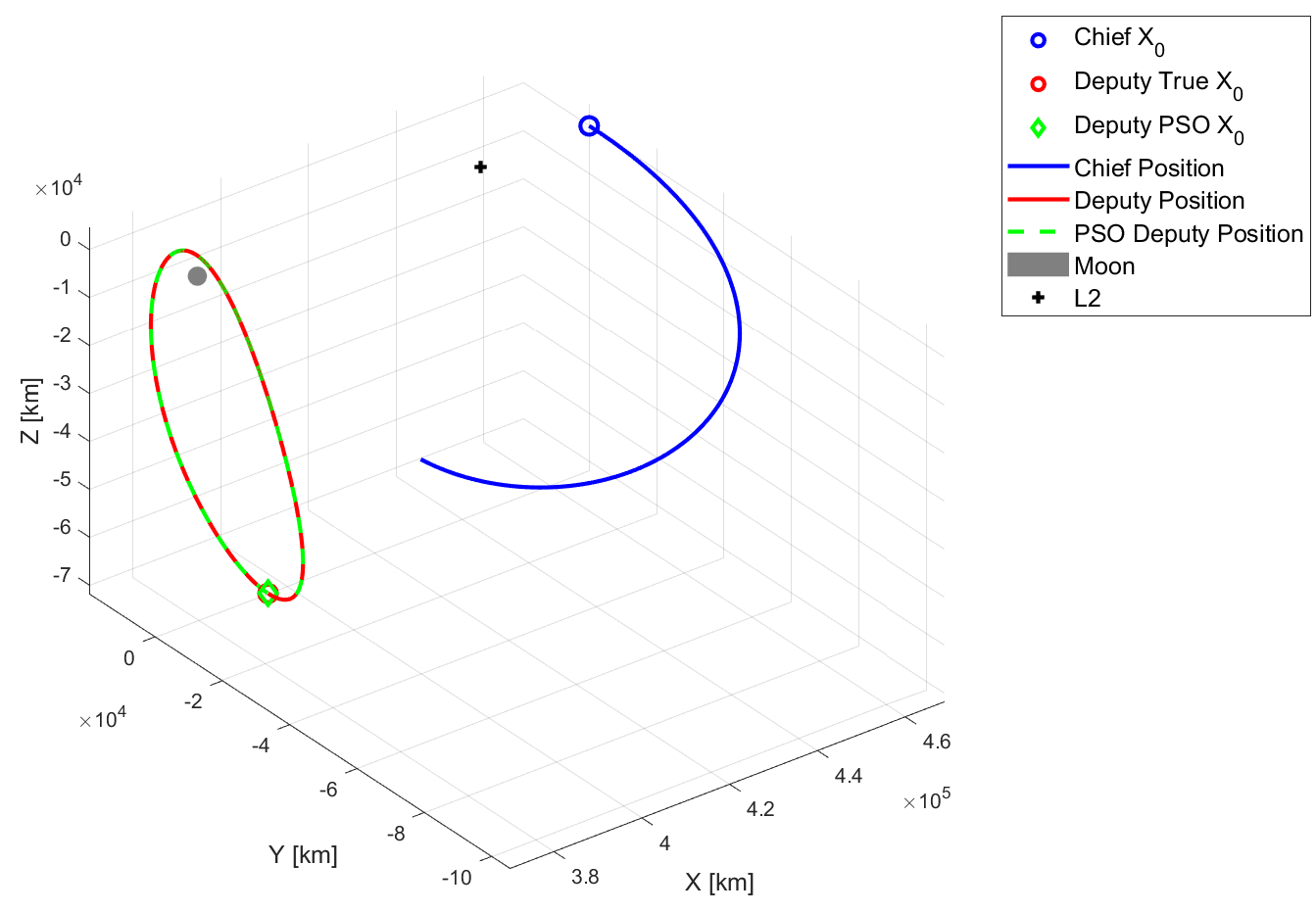}}
    \caption{Scenario 2: Chief and deputy HALO orbits depicted over 7-day period with PSO deputy orbit overlay.}
    \label{fig:scenario2:PSO:optimizer:results}
\end{figure}

Figure \ref{fig:scenario2:PSO:optimizer:results} shows the results of the simulation, with the propagated PSO solution for the deputy overlaid on the true deputy position. The chief's orbit is a wide, planar orbit, and the deputy is in a periodic HALO orbit. Note that the remaining figures for the second scenario are located in Appendix A for compactness. Overall the results are analogous to scenario 1. 

Figure \ref{fig:scenario2:minimum:cost} shows the global minimum cost as the PSO population moves towards the optimum solution. Note that at times the cost appears to only be moving downwards slowly, but with the logarithmic scale the progress is faster than it appears. Figure \ref{fig:scenario2:particle:positions} shows the initial and final particle distributions for the initial position of the deputy. Due to scale the global best and true solution at the center are difficult to see, but there is a collection of particles clustered on the truth. Figure \ref{fig:scenario2:particle:velocities} shows the initial and final distributions of velocity particles. Figure \ref{fig:scenario2:position:velocity:error} shows the error in initial position and velocity as the PSO converges to the truth. Velocities converge more quickly than the position states. Figure \ref{fig:scenario2:range:speed:error} shows the range and speed errors as the PSO converged.  

\section{Conclusions} 
Particle swarm optimization shows promise for providing IOD results for cislunar SDA applications. Scenarios tested were representative of possible SDA tasks in cislunar space. The goal being deriving a usable initial state guess for the deputy satellite to initiate proximity operations or enter a satellite into a catalog of cislunar objects for further tracking. PSO global optimization was able to get an initial state estimate that was close enough to seed a local optimizer. Further optimization through nonlinear least squares then found the initial state to much greater accuracy. Further improvement in the initial condition for PSO, and looking at varying weighting parameters is a subject of future investigation to improve the IOD results. Porting the code to use GPU vectorization in Python, and extension to the elliptical restricted three-body problem are also subjects of future work.

\section*{Acknowledgments}
This work was partially supported by the National Defense Science and Engineering Graduate (NDSEG) Fellowship program.

\clearpage
\appendix
\section*{Appendix A: Additional Figures}
\label{sec:appendix:A}

\begin{figure}[tbh!]
    \centerline{\includegraphics[scale=0.49]{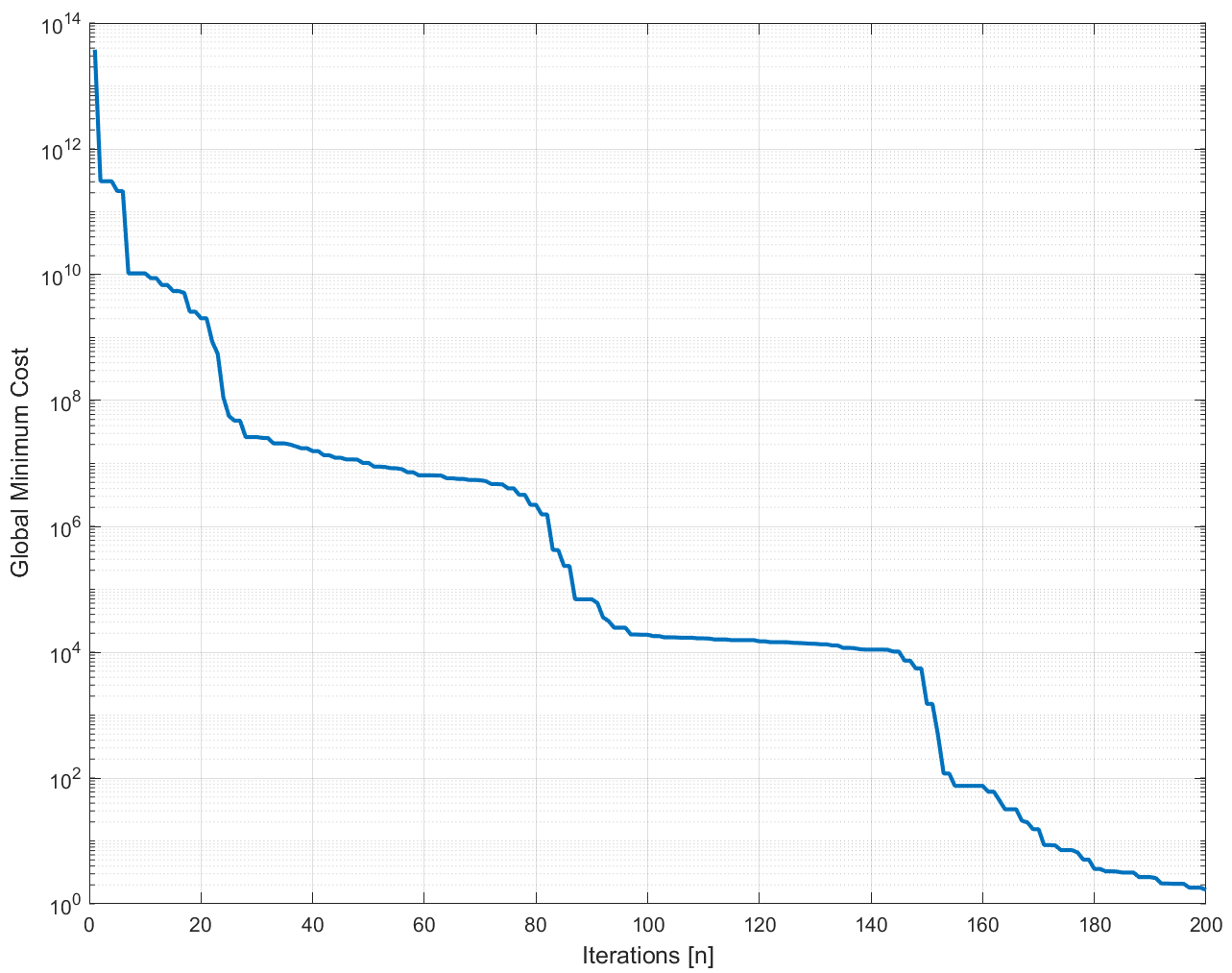}}
    \caption{Scenario 2: Global cost per iteration, decreasing until iteration counter is reached or tolerance is met.}
    \label{fig:scenario2:minimum:cost}
\end{figure}

\begin{figure}[tbh!]
    \centerline{\includegraphics[scale=0.49]{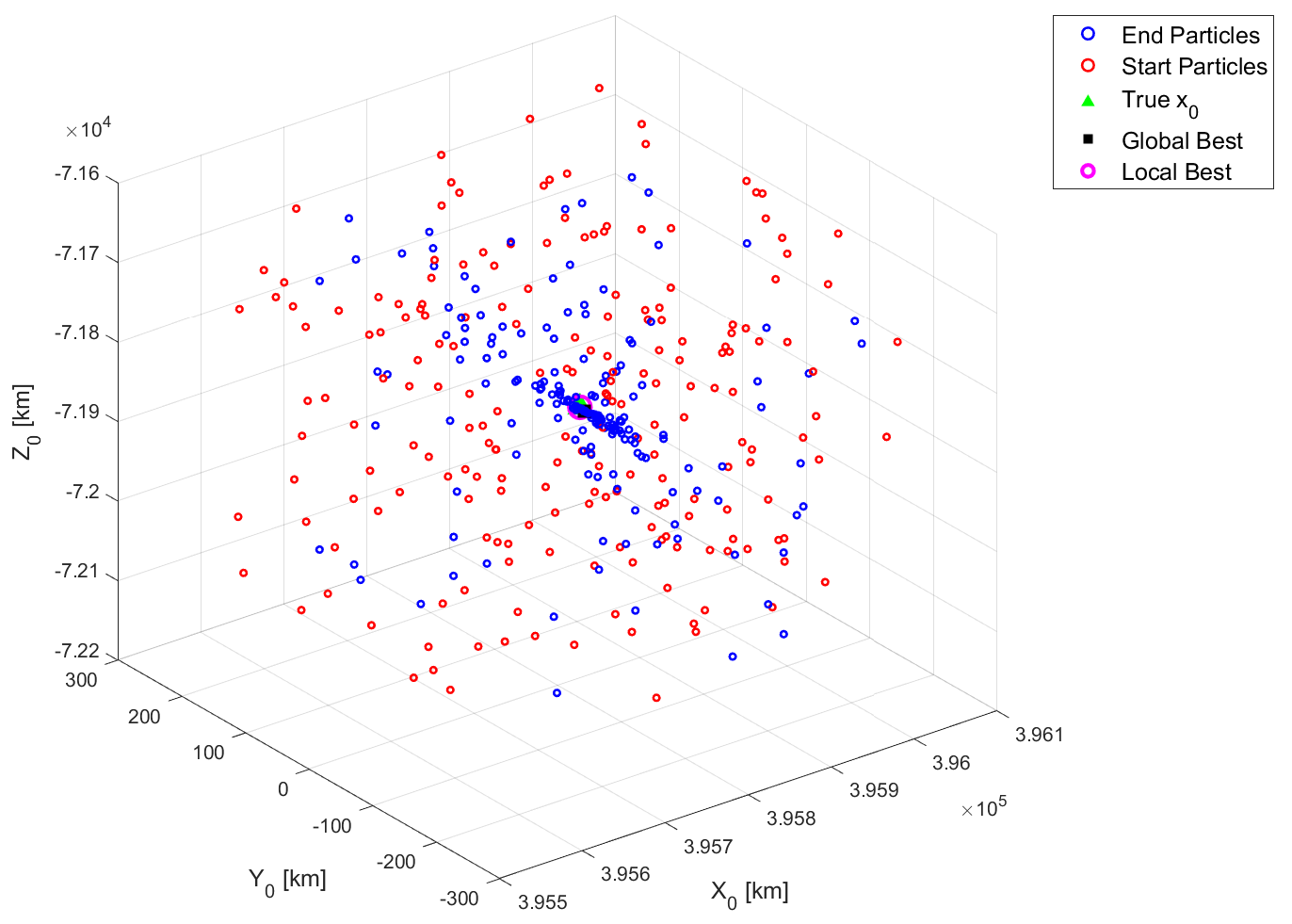}}
    \caption{Scenario 2: Initial (red) and final state (blue) particle positions in three axes, swarming to global best.}
    \label{fig:scenario2:particle:positions}
\end{figure}

\begin{figure}[tbh!]
    \centerline{\includegraphics[scale=0.49]{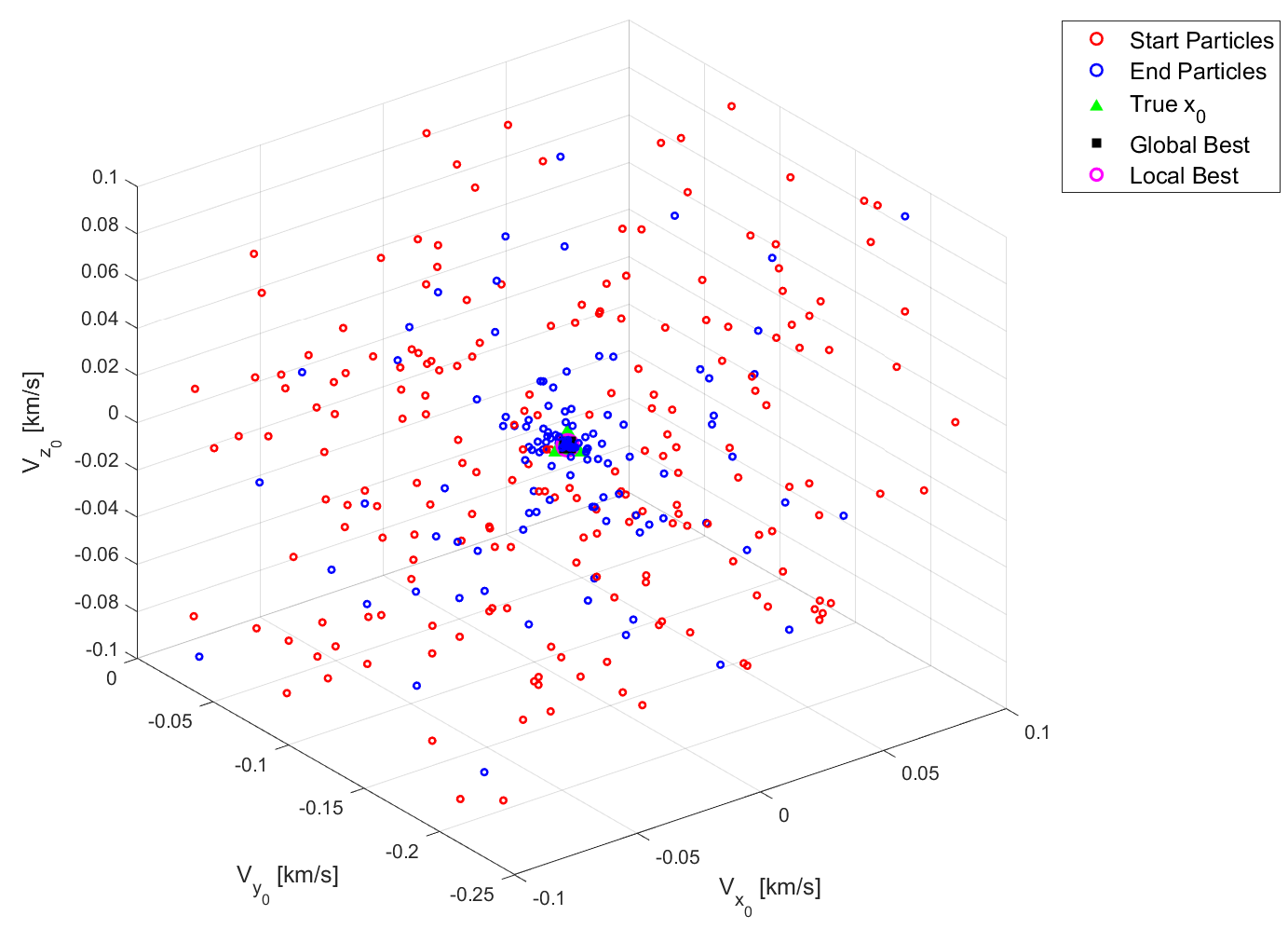}}
    \caption{Scenario 2: Initial (red) and final state (blue) of particle velocity in three axes.}
    \label{fig:scenario2:particle:velocities}
\end{figure}

\begin{figure}[tbh!]
    \centerline{\includegraphics[scale=0.49]{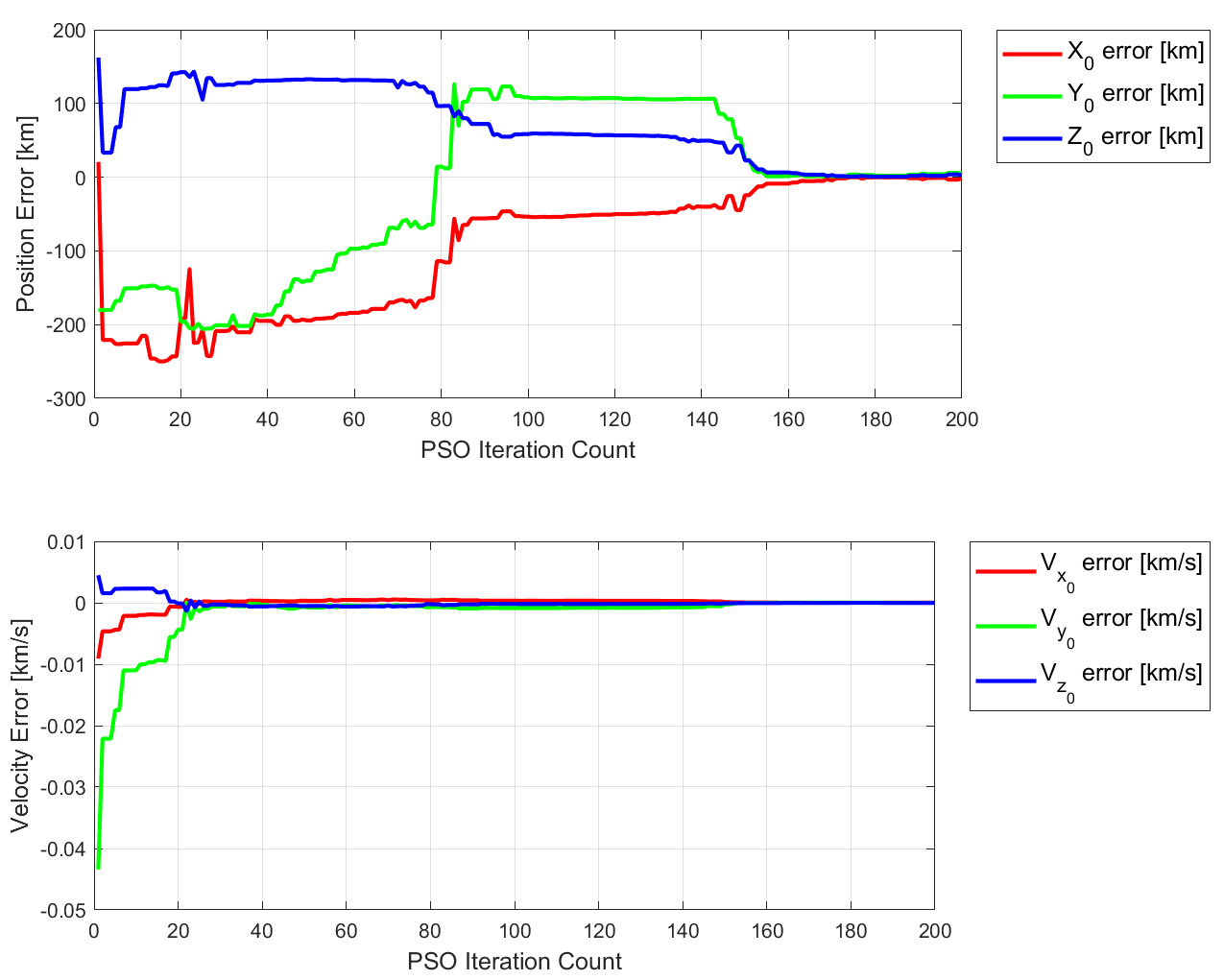}}
    \caption{Scenario 2: Position error in each axis per iteration (top) and velocity error in each axis per iteration (bottom).}
    \label{fig:scenario2:position:velocity:error}
\end{figure}
\clearpage
\begin{figure}[H]
    \centerline{\includegraphics[scale=0.49]{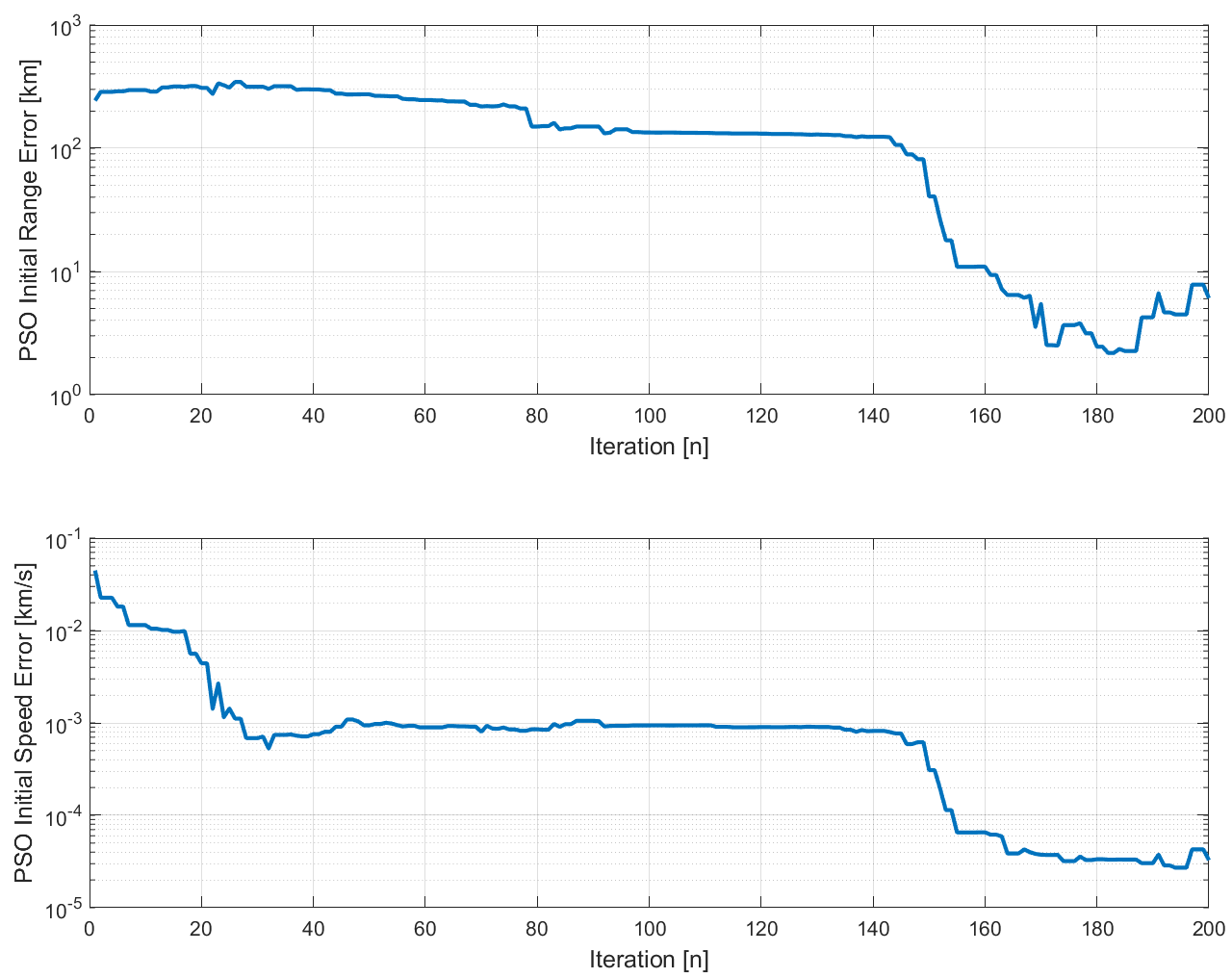}}
    \caption{Scenario 2: Magnitude of position error per iteration (top) and magnitude of velocity error per iteration (bottom).}
    \label{fig:scenario2:range:speed:error}
\end{figure}

\clearpage
\bibliographystyle{AAS_publication}
\newpage
\bibliography{aas_2022_refs}

\end{document}